\documentclass{elsart1p}
\usepackage{epsfig}
\usepackage{amsmath,mathrsfs,amsopn,bm,bbold}

\def\lsim{\raise0.3ex\hbox{$<$\kern-0.75em\raise-1.1ex\hbox{$\sim$}}}
\def\gsim{\raise0.3ex\hbox{$>$\kern-0.75em\raise-1.1ex\hbox{$\sim$}}}

\begin{document}

\begin{frontmatter}

\title{Fluctuations of Goldstone modes and the chiral transition in QCD}

\author{Frithjof Karsch (RBC-Bielefeld collaboration)}

\address{Physics Department, Brookhaven National Laboratory, Upton, NY 1973, USA\\
and Fakult\"at f\"ur Physik, Universit\"at Bielefeld, D-33615 Bielefeld, Germany}
\ead{karsch@bnl.gov}

\begin{abstract}
% Text of abstract
We provide evidence for the influence of thermal fluctuations of Goldstone 
modes on the chiral condensate at finite temperature. We show that at 
fixed temperature, $T < T_c$, in the vicinity of the chiral transition 
temperature this leads to a characteristic dependence of the chiral 
condensate on the square root of the light
quark mass ($m_l$), which is expected for 3-dimensional models with broken
O(N) symmetry. As a consequence the chiral susceptibility 
shows a strong quark mass dependence for all temperatures below $T_c$
and diverges like $1/\sqrt{m_l}$ in the chiral limit.
\end{abstract}

\end{frontmatter}

% main text
\section{Introduction}
\label{Intro}
Establishing the properties of the chiral phase transition in QCD is one
of the outstanding problems in lattice simulations of QCD thermodynamics.
Although several attempts have been undertaken to establish the 
unique scaling behavior in the limit of vanishing light quark masses
no clear-cut evidence for the expected $O(N)$ scaling,
related to the underlying chiral symmetry of the QCD Lagrangian, has 
been found. This may have several reasons related to common problems
in lattice calculations: cut-off effects, finite volume effects,
explicit flavor symmetry breaking and/or too large quark masses.

We will discuss here yet another effect that influences the scaling 
behavior of thermodynamic quantities in the vicinity of the finite
temperature chiral transition of QCD and may compete with the 
universal scaling in the vicinity of $T_c$. This is related to 
singularities in chiral observables, e.g. the chiral susceptibility,
that are induced at ANY temperature in the chiral symmetry broken
phase. The existence of these singularities induced by fluctuations
of Goldstone modes are well known from the analysis of 3 and 4 
dimensional statistical models and field theories with global $O(N$)
symmetries that are spontaneously broken at low temperature 
\cite{Wallace,Leutwyler,Hasenfratz,Smilga}.   

\section{Chiral condensate and susceptibility for \boldmath$T\lsim T_c$}

At vanishing temperature properties of QCD in the light quark sector
are described by an effective 4-d, $O(4)$ symmetric Lagrangian; chiral
symmetry is broken and, what is of interest here, at non-zero quark mass 
the light quark chiral condensate, $\langle \bar\psi \psi \rangle$, 
receives logarithmic corrections proportional to $m_l\ln m_l$ 
that are due to the presence of light Goldstone modes \cite{Leutwyler}. 
This leads to a logarithmic divergence of the chiral susceptibility, 
$\chi_m \sim {\rm d}\langle \bar\psi \psi \rangle / {\rm d} m$.
In three dimensions the analog to
these logarithmic terms are corrections that are proportional to the
square root of the quark mass \cite{Wallace}. As a consequence $\chi_m$ 
will diverge in the chiral limit like $\chi_m \sim 1/\sqrt{m_l}$. 

In the vicinity of the QCD phase transition a 3-d, $O(4)$ symmetric
effective theory is expected to describe the universal properties of 
the transition \cite{Pisarski} which will also control the properties 
of the chiral condensate for temperatures close to $T_c$. 
We thus expect that in the chiral limit $\chi_m$ will diverge for 
all values of the temperature $T\le T_c$,
\begin{equation}
\chi_m  \sim
\begin{cases}
m_l^{-1/2} &,\;  T \le T_c \\[1mm]
m_l^{1/\delta -1} &,\; T=T_c
\end{cases}
\label{scaling}
\end{equation}
where $1/\delta -1 \simeq 0.79$ for 3-d $O(4)$ as well as $O(2)$ symmetric
spin models.

\begin{figure}[t]
\begin{center}
%\begin{minipage}[c]{14.5cm%}
%\begin{center}
\epsfig{file=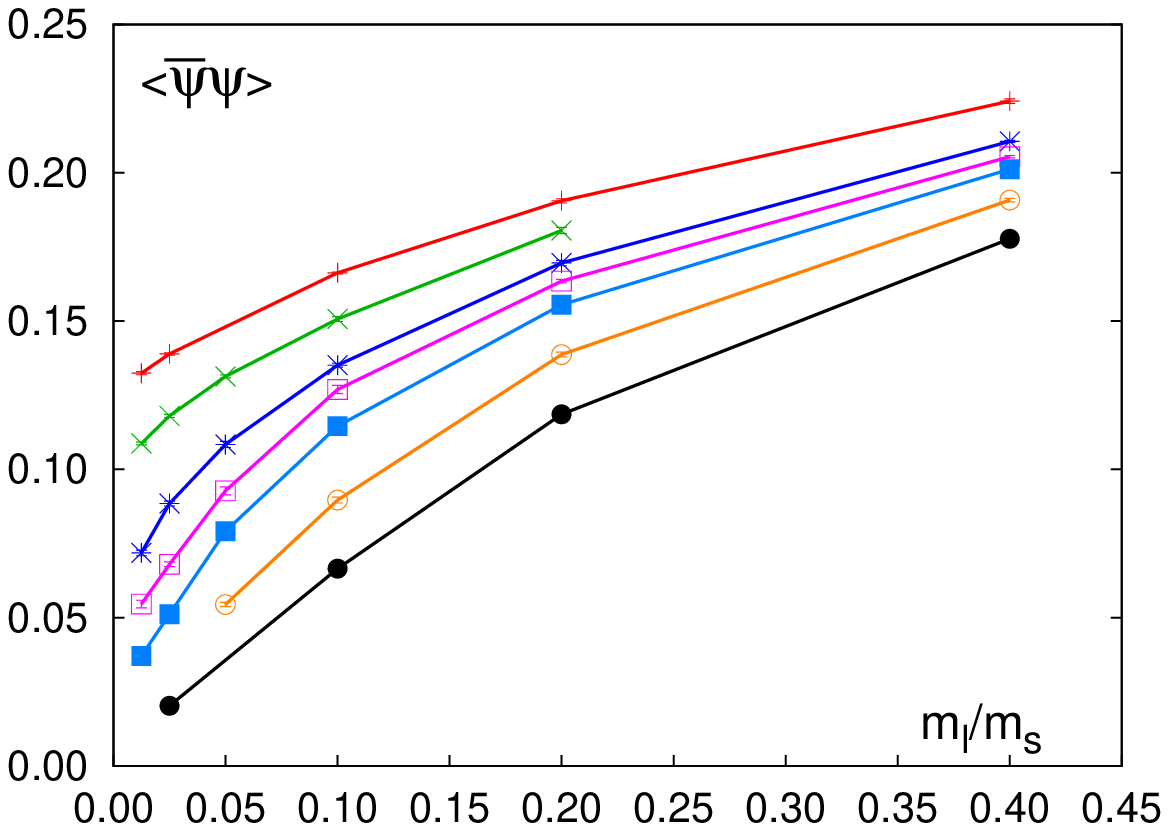, width=6.6cm}
\epsfig{file=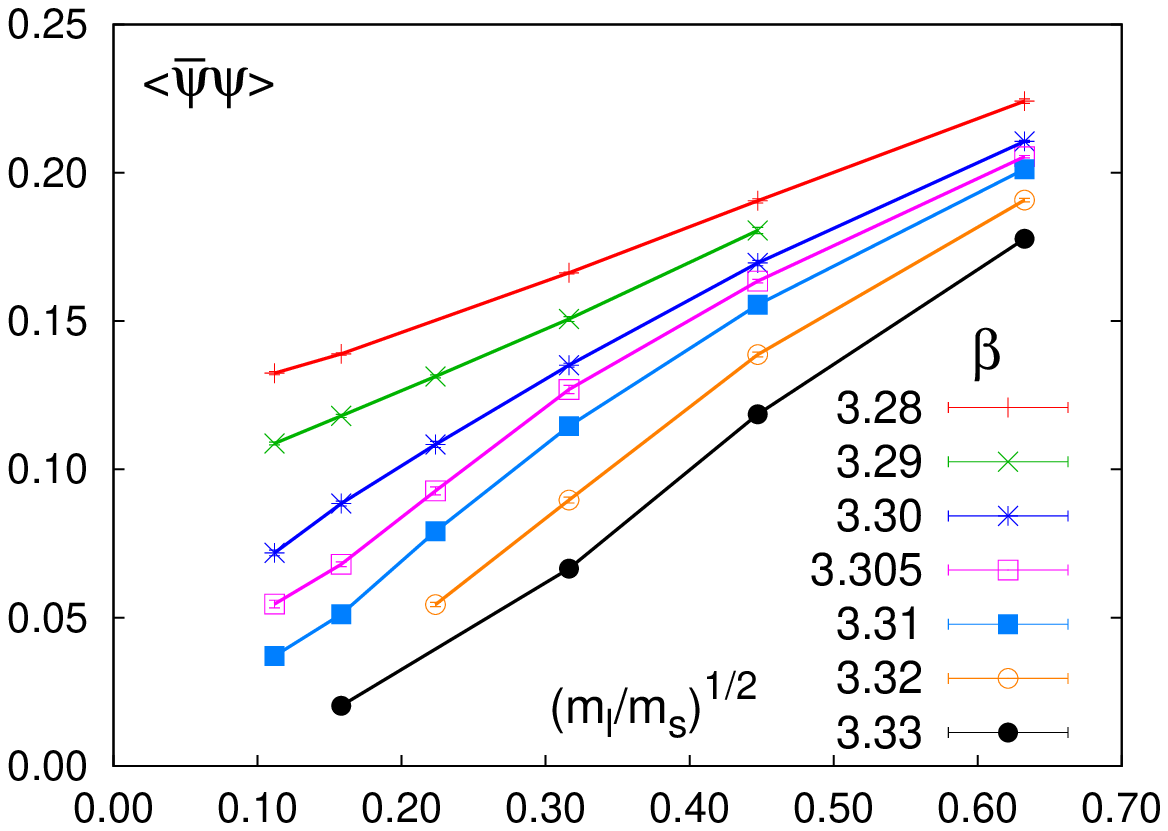, width=6.6cm}
%\end{center}
%\end{minipage}
\end{center}
\caption{Chiral condensate calculated on lattices of size
$N_\sigma^3\times4$ in $(2+1)$-flavor QCD for various 
values of the light quark mass ($m_l$) in units of the strange quark mass
($m_s$).  $N_\sigma$ from $32$ for the two lightest quark mass values 
to $8$ for the heaviest masses. For small quark masses the critical
coupling for the chiral transition temperature on this size lattices
has been determined to be in the range $\beta_c=3.305 - 3.31$ \cite{Cheng}.
}
\label{fig:pbp}
\end{figure}

The appearance of a square root singularity has, in fact, been 
established in the vicinity of the chiral phase transition of an 
$SU(3)$ gauge theory coupled to adjoint fermions \cite{Karsch} 
where it is apparent already for moderately small values of the 
quark mass. Similar evidence for the role of light Goldstone modes 
in QCD, however, has been missing so far.  

In Fig.~\ref{fig:pbp} we show the light quark chiral condensate, 
$\langle \bar\psi \psi \rangle$
calculated in (2+1)-flavor QCD with improved staggered fermions (p4-action)
on lattices of
size $N_\sigma^3\times 4$ for various values of the gauge coupling. 
The figures show results for $\langle \bar\psi \psi \rangle$,  plotted 
versus the light quark mass ($m_l$) expressed in units of the strange 
quark mass ($m_s$) (left) and the square root of this ratio (right). 
The right hand figure indeed suggests that
the strong curvature of $\langle \bar\psi \psi \rangle$ as function of
$m_l$ is well accounted for by a $\sqrt{m_l}$ dependence for 
$T < T_c$. In any case, the weaker than linear dependence
of $\langle \bar\psi \psi \rangle$ on $m_l$ leads to a strong
quark mass dependence of the chiral susceptibility, $\chi_m$. 
This susceptibility receives two contributions, usually referred to as 
disconnected and connected parts, 
$\chi_m = \chi_{dis} + \chi_{con}$, with
\begin{eqnarray} 
\hspace*{-0.5cm}\frac{\chi_{dis}}{T^2} = \frac{N_\tau}{16N_\sigma^3}\biggl( 
\langle ({\rm Tr} D_l^{-1})^2\rangle - \langle {\rm Tr} D_l^{-1}\rangle 
\biggr) \; &,&\; 
\frac{\chi_{con}}{T^2} = -\frac{N_\tau^2}{4}\sum_{x} \langle D_l^{-1}(x,0)
D_l^{-1}(0, x)\rangle \; .
\nonumber
\label{chim}
\end{eqnarray}
Here $D_l$ denotes the staggered fermion matrix for light quarks.
In general, both contributions are sensitive to the 
Goldstone modes and will receive divergent contributions,
$\chi_{dis,con}\sim 1/\sqrt{m_l}$ \cite{Smilga}. 
We will restrict our analysis here to the disconnected part.

\begin{figure}[t]
\begin{center}
%\begin{minipage}[c]{14.5cm%}
%\begin{center}
\epsfig{file=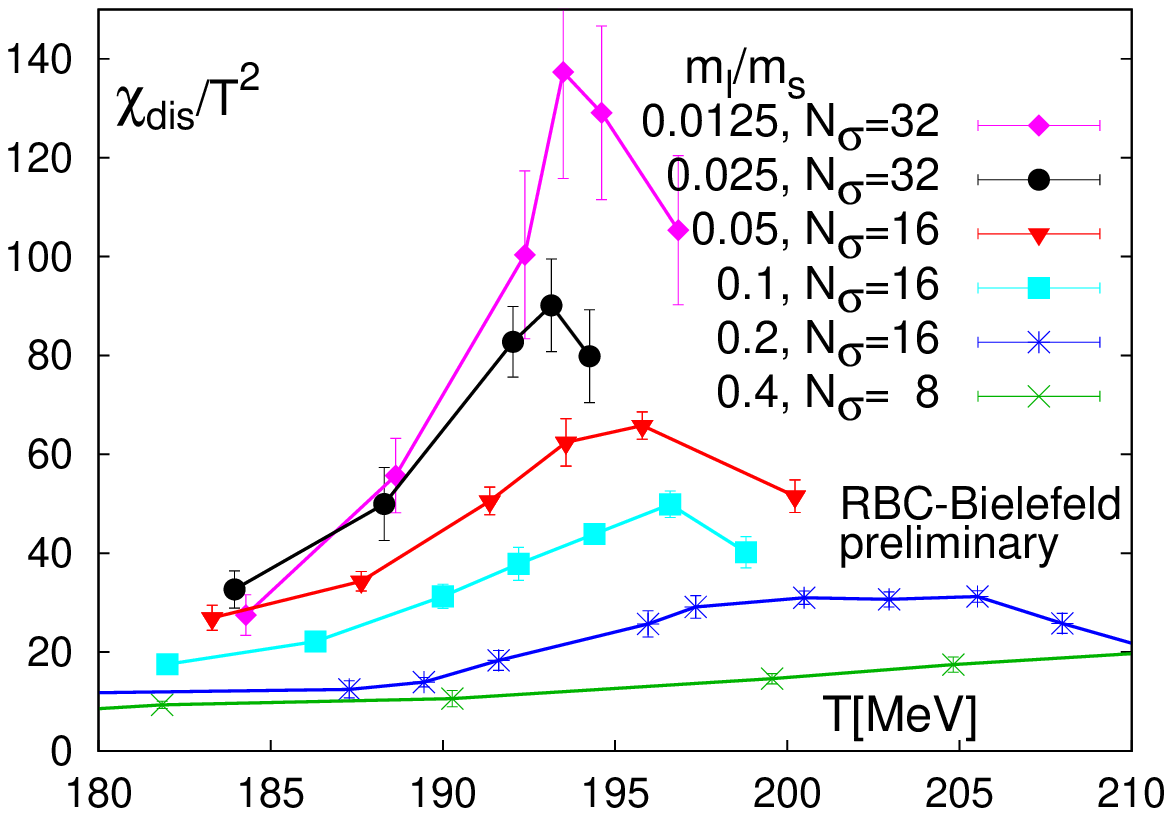, width=6.6cm}
\epsfig{file=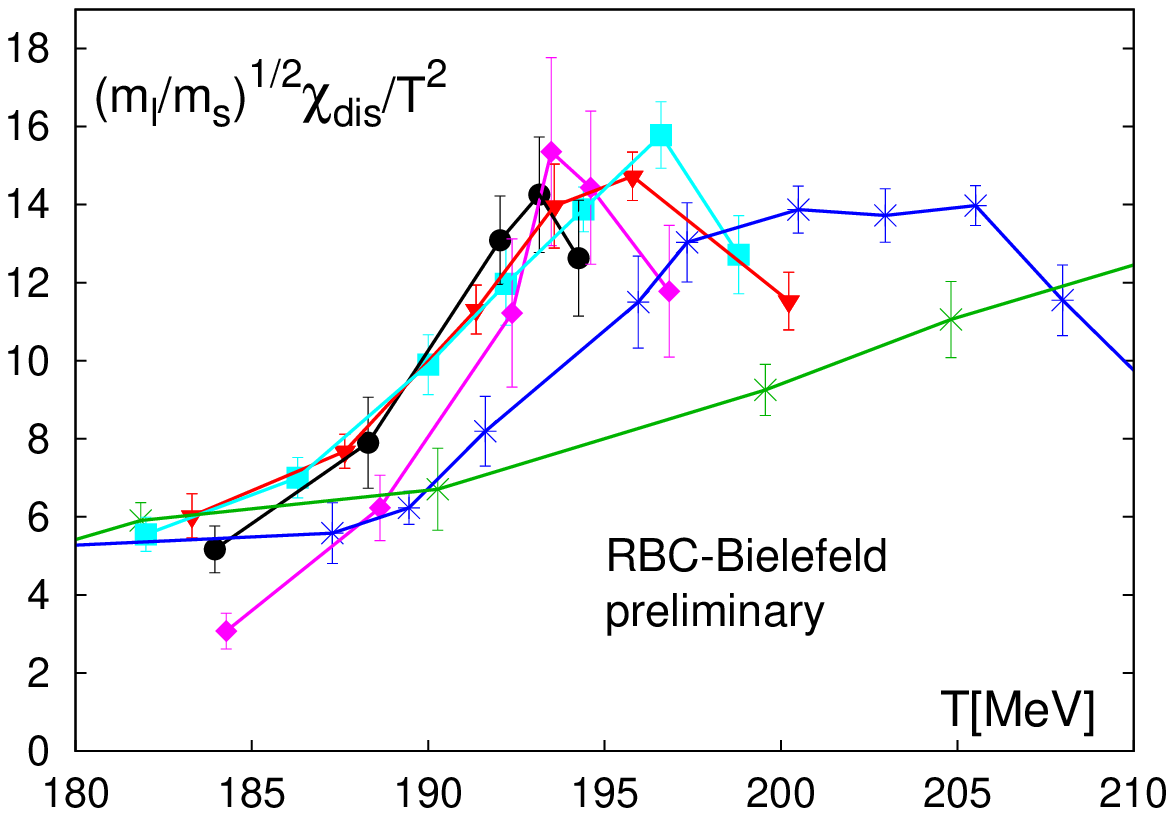, width=6.6cm}
%\end{center}
%\end{minipage}
\end{center}
\caption{Disconnected part of the chiral susceptibility calculated on 
lattices of size $N_\sigma^3\times 4$ in $(2+1)$-flavor QCD for various
values of the light quark mass. 
}
\label{fig:chi}
\end{figure}

In Fig.~\ref{fig:chi} we show $\chi_{dis}/T^2$ calculated for various 
values of $m_l$. In these calculations the strange quark mass
is close to its physical value and $m_l$ reaches values 
well below the physical one; a ratio $m_l/m_s \simeq 0.04$ corresponds
in the continuum limit to the physical quark mass ratio.
The strong quark mass dependence in a wide 
temperature interval is obvious from Fig.~\ref{fig:chi}(left); its dominant
contribution in the low temperature regime is indeed well
described by $\chi_m \sim 1/\sqrt{m_l}$ as can be seen from the rescaled
data shown in Fig.~\ref{fig:chi}(right). We note that this seems to hold 
for all quark masses, $m_l\le 0.1 m_s$, although at present our data sample
for the lightest quark mass is not yet good enough to substantiate this
statement also for this light mass value, which corresponds to a light
pseudo-scalar mass of about $70$~MeV. We furthermore, note that the
square root behavior also dominates the quark mass dependence in the peak
region, {\it i.e.} there is no evidence for a stronger singularity,
which is expected to arise from the singular part of the partition 
function of 3-d, O(4) (or O(2)) symmetric models, $\chi_m^{singular}\sim
m_l^{-0.79}$.
 
\section{Towards the continuum limit of (2+1)-flavor QCD}

\begin{figure}[t]
\begin{center}
%\begin{minipage}[c]{14.5cm%}
%\begin{center}
\epsfig{file=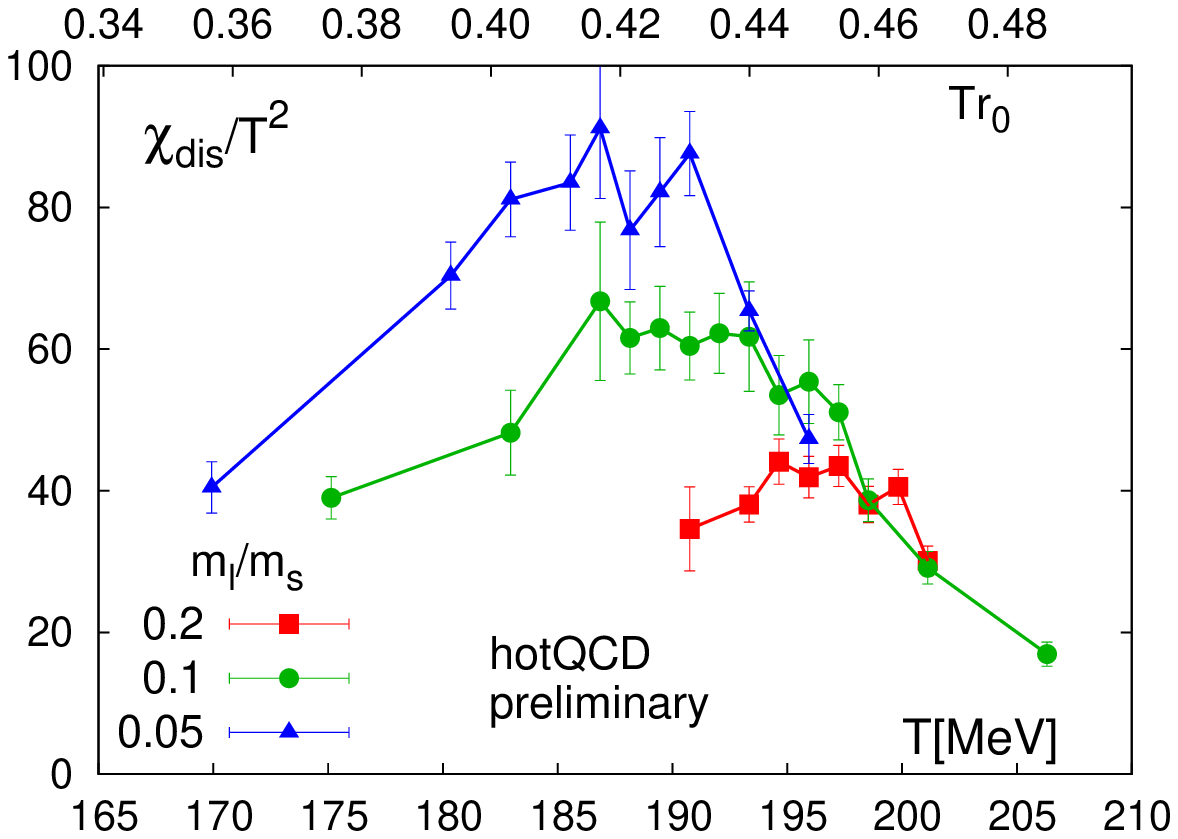, width=6.6cm}
\epsfig{file=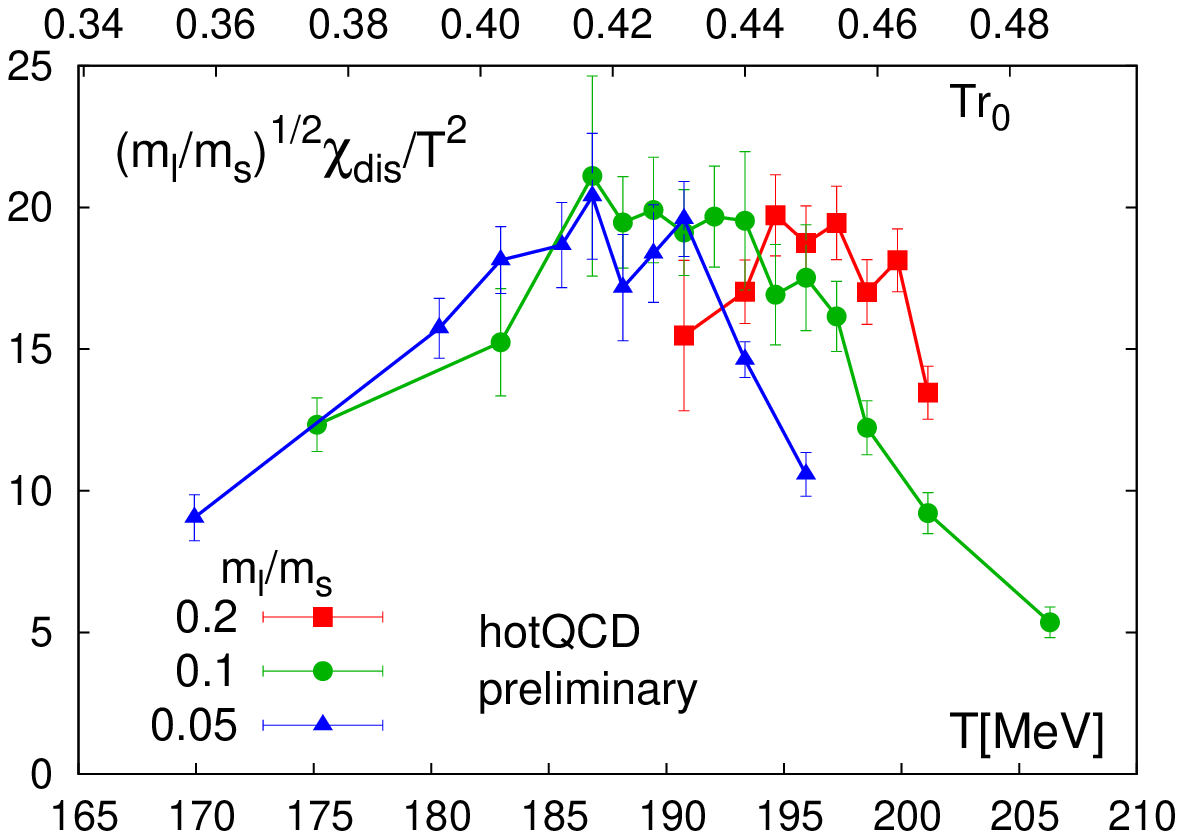, width=6.6cm}
%\end{center}
%\end{minipage}
\end{center}
\caption{Disconnected part of the chiral susceptibility calculated on 
lattices of size $32^3\times8$ in $(2+1)$-flavor QCD for 
$m_l/m_s = 0.05,\; 0.1$ and $0.2$, respectively.
}
\label{fig:chi8}
\end{figure}

The general features of the quark mass dependence of the disconnected 
part of the chiral susceptibility are also apparent in calculations
currently performed by the hotQCD collaboration on lattice with temporal 
extent $N_\tau=8$ \cite{deTar}, {\it i.e.} closer to the continuum limit. In 
Fig.~\ref{fig:chi8} we show $\chi_{dis}/T^2$ and the rescaled 
susceptibility. The left hand part of the figure shows that for 
temperatures above the transition region the susceptibilities 
quickly become quark mass independent. 
This is expected as the chiral condensates will be linear
in the quark mass in the chirally symmetric phase of QCD. Below the 
transition region we again observe a strong quark mass dependence that 
is consistent with $\sim 1/\sqrt{m_l}$. 

The weak quark mass dependence of $\chi_m$ in the high temperature 
regime suggest that the susceptibility can already
be described by the limiting form for the $m_l\rightarrow 0$ limit,
\begin{equation}
\chi_m   = 
\begin{cases}
\infty &,\;  T \le T_c \\[1mm]
A \left(\frac{T-T_c}{T_c}\right)^{-\gamma} &,\; T > T_c
\end{cases}
\label{chiral}
\end{equation}
A consistent picture for the chiral susceptibility thus seems to emerge
despite the missing evidence for $O(N)$ scaling for its peak height. 

\section*{Acknowledgments}
The numerical calculations presented here have been carried out on
the QCDOC supercomputers of the RIKEN-BNL Research Center, the  
BlueGene/L computers at the New York Center for Computational Sciences
and the LLNL. The work of FK has been supported by DOE
%the U.S. Department of Energy 
under Contract No. DE-AC02-98CH10886.


\begin{thebibliography}{00}

\bibitem{Wallace}
D.J.Wallace and R.K.P. Zia, Phys. Rev. {\bf B12} (1975) 5340
\bibitem{Leutwyler}
J. Gasser and H. Leutwyler, Phys. Lett. {\bf B184} (1987) 83
\bibitem{Hasenfratz} 
P. Hasenfratz and H. Leutwyler, Nucl. Phys. {\bf B343} (1990) 241
\bibitem{Smilga}
A. V. Smilga, Phys. Lett. {\bf B318} (1993) 531; \\
A. V. Smilga and J. J. M. Verbaarschot, Phys. Rev. {\bf D54} (1996) 1087
\bibitem{Pisarski}
R. Pisarski and F. Wilczek, Phys. Rev. {\bf D29} (1984) 338
\bibitem{Karsch}
F. Karsch and M. L\"utgemeier,  Nucl. Phys. {\bf B550} (1999) 449;\\
J. Engels, S. Holtmann and T. Schulze, Nucl. Phys. {\bf B724} (2005) 357 
\bibitem{Cheng}
M. Cheng et al. (RBC-Bielefeld), Phys. Rev. {\bf D74} (2006) 054507
\bibitem{deTar}
C. DeTar, PoS 001 (LAT2008) \\ 
R.~Gupta  [for the HotQCD Collaboration], PoS 170 (LAT2008),
%``The QCD EoS from simulations on BlueGene L Supercomputers at LLNL and
%NYBlue,''
arXiv:0810.1764 [hep-lat].
\end{thebibliography}
\end{document}